\documentclass[twocolumn]{aastex63}

\usepackage{epsfig,natbib}
\usepackage{graphicx}
\usepackage{gensymb}
\usepackage{amsmath}
\usepackage{longtable}
\usepackage{xspace}
\usepackage{hyperref}
\usepackage{lipsum}
\usepackage{physics}
\newcommand{\mse}{\mbox{m\,s$^{-1}$}}

\newcommand{\msun}{M$_{\odot}~$}
\newcommand{\msune}{M$_{\odot}$}

\newcommand{\rsune}{R$_{\odot}$}
\newcommand{\lsun}{L$_{\odot}~$}
\newcommand{\lsune}{L$_{\odot}$}

\newcommand{\mearth}{$M_\earth$~}

\newcommand{\rearth}{$R_\earth$~}

\newcommand{\mstar}{\ensuremath{M_{\star}}}
\newcommand{\rstar}{\ensuremath{R_{\star}}}
\newcommand{\lstar}{\ensuremath{L_{\star}}}

\newcommand{\rphk}{\ensuremath{R'_{\mbox{\scriptsize HK}}}}
\newcommand{\shk}{\ensuremath{S_{\mbox{\scriptsize HK}}}}
\newcommand{\lrphk}{\ensuremath{\log{\rphk}}}

\newcommand{\bjdtdb}{\ensuremath{\rm {BJD_{TDB}}}}

\newcommand{\name}{Gl 414A\ }
\newcommand{\ename}{Gl 414A}

\newcommand{\nobskeck}{126\ }
\newcommand{\kecksdate}{Jan 14, 1997\ }
\newcommand{\keckedate}{June 16, 2019\ }

\newcommand{\nobsapf}{351\ }
\newcommand{\apfsdate}{Oct 23, 2013\ }
\newcommand{\apfedate}{June 3, 2019\ }

\newcommand{\msinic}{$56.27 \substack{ +10.43 \\ -9.91}$\ }
\newcommand{\ac}{1.43 \pm{0.06}\ }
\newcommand{\Tc}{123.3  \pm 13.2\ }
\newcommand{\rc}{$8.78 \substack{+ 4.03 \\ - 2.66}$\ }
\newcommand{\msinib}{$8.78 \substack{ +3.10 \\ -2.47}$\ }
\newcommand{\ab}{0.240\ \pm 0.01}
\newcommand{\Tb}{303.7 \pm 32.5\ }
\newcommand{\rb}{$2.95 \substack{+ 1.11 \\ - 0.91}$\ }

\shortauthors{Dedrick et al.}
\shorttitle{\name}
\begin{document}

\title{Two Planets Straddling the Habitable Zone of The Nearby K dwarf \name}

\author[0000-0001-9408-8848]{Cayla M.\ Dedrick}
\affiliation{California Institute of Technology, Pasadena, CA 91125, USA}
\affiliation{Department of Astronomy \& Astrophysics, The Pennsylvania State University, 525 Davey Lab, University Park, PA 16802, USA}

\author[0000-0003-3504-5316]{Benjamin J.\ Fulton}
\affiliation{California Institute of Technology, Pasadena, CA 91125, USA}
\affiliation{IPAC-NASA Exoplanet Science Institute, Pasadena, CA 91125, USA}

\author[0000-0002-0822-3095]{Heather A.\ Knutson}
\affiliation{California Institute of Technology, Pasadena, CA 91125, USA}

\author[0000-0001-8638-0320]{Andrew W.\ Howard}
\affiliation{California Institute of Technology, Pasadena, CA 91125, USA}

\author[0000-0002-9539-4203]{Thomas G. Beatty}
\affiliation{Department of Astronomy and Steward Observatory, University of Arizona, Tucson, AZ 85721}

\author[0000-0002-1617-8917]{Phillip A.\ Cargile}
\affiliation{Center for Astrophysics $\vert$ Harvard \& Smithsonian, 60 Garden Street, Cambridge, MA 02138, USA}

\author[0000-0003-0395-9869]{B.\ Scott Gaudi}
\affiliation{Department of Astronomy, The Ohio State University, 140. W. 18th Ave., Columbus, OH 43210, USA}

\author[0000-0001-8058-7443]{Lea A.\ Hirsch}
\affiliation{Kavli Institute for Particle Astrophysics and Cosmology, Stanford University, Stanford, CA 94305, USA}

\author[0000-0002-4236-9020]{Rudolf B. Kuhn}
\affiliation{South African Astronomical Observatory, PO Box 9, Observatory, 7935, Cape Town, South Africa}
\affiliation{Southern African Large Telescope, PO Box 9, Observatory, 7935, Cape Town, South Africa}
\author[0000-0003-2527-1598]{Michael B. Lund} 
\affiliation{IPAC-NASA Exoplanet Science Institute, Pasadena, CA 91125, USA}

\author[0000-0001-5160-4486]{David J. James}
\affiliation{Center for Astrophysics $\vert$ Harvard \& Smithsonian, 60 Garden Street, Cambridge, MA 02138, USA}
\affiliation{Black Hole Initiative at Harvard University, 20 Garden Street, Cambridge, MA 02138, USA}

\author[0000-0002-6115-4359]{Molly R.\ Kosiarek}
\altaffiliation{NSF Graduate Research Fellow}
\affiliation{Department of Astronomy and Astrophysics, University of California, Santa Cruz, CA 95064, USA}

\author[0000-0002-3827-8417]{Joshua Pepper}
\affiliation{Department of Physics, Lehigh University, 16 Memorial Drive East, Bethlehem, PA, 18015, USA}

\author[0000-0003-0967-2893]{Erik A.\ Petigura}
\affiliation{Department of Physics and Astronomy, University of California, Los Angeles, CA 90095, USA}

\author[0000-0001-8812-0565]{Joseph E. Rodriguez}
\affiliation{Center for Astrophysics $\vert$ Harvard \& Smithsonian, 60 Garden Street, Cambridge, MA 02138, USA}

\author[0000-0002-3481-9052]{Keivan G.\ Stassun}
\affiliation{Vanderbilt University, Department of Physics \& Astronomy, 6301 Stevenson Center Ln., Nashville, TN 37235, USA}
\affiliation{Fisk University, Department of Physics, 1000 18th Ave. N., Nashville, TN 37208, USA}

\author[0000-0002-5951-8328]{Daniel J. Stevens}
\altaffiliation{Eberly Research Fellow}
\affiliation{Department of Astronomy \& Astrophysics, The Pennsylvania State University, University Park, PA 16802, USA}
\affiliation{Center for Exoplanets and Habitable Worlds, The Pennsylvania State University, 525 Davey Lab, University Park, PA 16802, USA}

\begin{abstract}
    We present the discovery of two planets orbiting the nearby (D=11.9 pc) K7 dwarf Gl 414A. Gl 414A b is a sub-Neptune mass planet with M$_b \sin{i_b} = 9.28^{+3.19}_{-2.54}$ M$_\oplus$ and a semi-major axis of 0.24 $\pm$ 0.01 au. Gl 414A c is a sub-Saturn mass planet with $M_c \sin{i_c} = 59.48^{+9.98}_{-9.69}$ M$_\oplus$ and a semi-major axis of 1.43 $\pm$ 0.06 au. We jointly analyzed radial velocity data from Keck/HIRES and the Automated Planet Finder at Lick Observatory, as well as photometric data from KELT, to detect the two planets and two additional signals related to the rotationally-modulated activity and the long term magnetic activity cycle of the star. The outer planet in this system may be a potential candidate for future direct imaging missions.
\end{abstract}

\section{Introduction}
\label{sec:intro}

High precision radial velocity (RV) instruments have improved in performance in recent years as the community moves toward the $\sim$cm s$^{-1}$ precision range. When combined with high cadence observational campaigns, these measurements make it possible to discover new low mass planets around nearby stars. Statistical methods have also improved greatly, with techniques like Gaussian processes (GPs) to model activity signals \citep[e.g.,][]{Dai17} making it easier to differentiate between exoplanetary and low-level activity-related signals \citep{Bastien14}. The use of a GP ensures that our posteriors for the orbital parameters of the two planets properly account for the RV variations caused by rotationally-modulated star spots and other sources of stellar activity.

 The Eta-Earth Survey utilized the W.\ M.\ Keck Observatory and the HIRES instrument \citep{Vogt94} to search 230 of the nearest G, K, and M dwarf stars for low mass (3-30 $M_\Earth$) planets \citep{Howard09a}. This survey unveiled the mass-distribution of small planets and demonstrated that the occurrence rate of planets is a strong inverse function of their mass \citep{Howard10}. Our group continues to measure RVs for all of the Eta-Earth stars at reduced cadence with both Keck-HIRES and the Automated Planet Finder telescope \citep[APF,][]{Vogt14}. Some of these stars have been observed for more than two decades, providing sensitivity to planets with orbital periods of many years. At the same time, recent high cadence observing campaigns with the APF have been very important in detecting lower mass planets and disentangling planetary signals from stellar activity \citep{Fulton16,Fulton17b}.
 
 The Eta-Earth Survey stars are an ideal test set for utilizing Gaussian processes to model stellar activity in exoplanet detection because of the multi-decadal baseline and large number of measurements. We continue to observe stars from the Eta-Earth survey in hopes of finding more small planets with longer orbital periods. Exoplanets with orbital periods greater than 700 days make up less than 10\% of the confirmed planet sample (311 out of 4201)\footnote{\url{https://exoplanetarchive.ipac.caltech.edu/}}. Due to sensitivity constraints, a majority of these long-period planets are gas giants. 

In this paper, we report the discovery of two planets orbiting the nearby K dwarf \ename. This star shows RV variations caused by stellar activity with amplitudes comparable to the planetary signals, which makes it a perfect candidate for combining RV and photometry data in order to differentiate between the types of signals. In Section \ref{sec:star}, we provide an updated analysis of the properties of the host star. We describe our Doppler measurements from Keck/HIRES and APF/Levy in Section \ref{sec:rv}, and describe our photometry measurements from KELT in Section \ref{sec:phot}. In Section \ref{sec:analysis} we describe our joint analysis of the RV data and photometric data. This includes a Keplerian analysis of significant periodic signals in the RVs and a characterization using GPs to model stellar activity. Section \ref{sec:discussion} discusses the orbital dynamics, transit and direct imaging prospects. We summarize and conclude in Section \ref{sec:summary}.

\section{Stellar Properties}
\label{sec:star}

\ename, also known as HD 97101A and HIP 54646A, is a bright nearby K7V dwarf star \citep{Gray03} at a distance of $11.89 \pm 0.07$ pc \citep[][Gaia18 here-after]{Gaia18}. \name has an M2V dwarf companion \citep{Stephenson86} with a V-band magnitude of $9.98$ \citep{Zacharias12}. The projected on-sky separation between \name and Gl 414B is $34.34 \arcsec$ \citep{Mason01}, corresponding to a projected physical separation of 408 AU. The stars are widely enough separated that spectral contamination from Gl 414B is not a concern. 

\subsection{SpecMatch-Emp \& isoclassify} \label{SI}

We inferred stellar properties using the publicly available software package \texttt{SpecMatch-Emp} \citep{Yee17} to quantitatively compare our iodine-free template spectrum of \name to a suite of library HIRES spectra for stars with well-measured parameters. The \texttt{SpecMatch-Emp} analysis returned T$_{\rm eff} = 4120 \pm 70$ K, [Fe/H] $= 0.24 \pm 0.09$, and $\rstar  = 0.680 \pm 0.10$ \rsune.

Since \texttt{SpecMatch-Emp} does not return a stellar mass, we used the \texttt{isoclassify} package \citep{Berger20, Huber17} to place the star onto MESA Isochrones \& Stellar Tracks \citep[MIST]{Dotter16, Choi16} isochrone tracks. We followed the methodology of \citet{Fulton18} to derive physical stellar parameters from the spectroscopic parameters. We ran \texttt{isoclassify} in both ``direct'' mode and ``grid'' mode. In direct mode \texttt{isoclassify} relies on a single apparent magnitude ($m_K$ in this case), the extinction in that band, the distance modulus, and a bolometric correction obtained by interpolating the MIST/C3K model grid (Conroy et al., in prep). In grid mode, the star is compared to the isochrones using observational constraints. We informed the \texttt{isoclassify} fit by placing priors on Teff and [Fe/H] from our spectroscopic analysis, the Gaia parallax ($\pi = 84.08 \pm 0.471$ mas, \cite{Gaia18}), and the K-band apparent magnitude ($m_K = 4.979 \pm 0.018$ mag, \cite{Cutri03}). \texttt{isoclassify} then calculated the posterior probability density for $\rstar$ using the Stefan-Boltzmann law. In direct mode, we found that $\lstar = 0.119 \pm 0.005$ \lsun and $\rstar = 0.679 \pm 0.027$ \rsune. Running \texttt{isoclassify} in grid mode also allowed us to extract $\mstar= 0.650 \pm 0.028$ \msun and age = $11.2 \pm 5.9$ Gyr.

We carried out the same analysis to determine stellar properties of the companion, Gl 414B. \texttt{SpecMatch-Emp} returned T$_{\rm eff} = 3663 \pm 70$ K, [Fe/H] $= 0.08 \pm 0.09$, and $\rstar = 0.515 \pm 0.100$ \rsune. Because Gl 414B is sufficiently cool, \texttt{isoclassify} is able to determine masses in direct mode using empirical relations from \citep{Mann19}. We found that $\rstar = 0.548 \pm 0.017$ \rsune, $\mstar = 0.542 \pm 0.022$ \msune, and $\lstar = 0.048 \pm 0.005$ \lsune. In grid mode, we derive age = $12.4 \pm 5.2$ Gyr.

If the two stars formed in the same birth cloud, we would expect the value of [Fe/H] and the ages of the two stars to be the same, and indeed the values we derive from \texttt{SpecMatch-Emp} and \texttt{isoclassify} analysis are consistent to 1$\sigma$.

We adopt the values of T$_{eff}$ and [Fe/H] from \texttt{SpecMatch-Emp}, and the values of $\rstar$, $\mstar$, $\lstar$, and age from \texttt{isoclassify}. See Table \ref{tab:MS} for the final adopted stellar parameters for both stellar components in the system.

\subsection{MINESweeper} \label{MS}

We also determine stellar parameters using \texttt{MINESweeper}, a tool to model stellar photometry using isochrone priors.  Full details and validation of this technique can be found in \citet{Cargile2019}, but briefly, the program can fit broadband photometric stellar spectral energy distributions (SEDs) with models drawn from the MIST \citep{Choi2016} stellar isochrones. The SED models (and corresponding predicted photometry) are computed from grids of \texttt{ATLAS12} model atmospheres \citep{Kurucz1970} and the spectrum synthesis code \texttt{SYNTHE} \citep{Kurucz1993}. Both atmospheres and SEDs are computed in 1D assuming plane-parallel geometry and LTE. We adopt the solar abundances from \citet{Asplund2009}, which is also the abundance scale used in the MIST isochrones. Atomic and molecular line lists are adopted from the latest compilation of R. Kurucz (private communication), and have been astrophysically calibrated against ultra high resolution (R $>$ 200k) spectra of the Sun and Arcturus using the same model assumptions as adopted herein (Cargile et al., in prep.). The fit is performed using the nested sampling code \texttt{dynesty} \citep{Speagle2019}. 

We fit \name with \texttt{MINESweeper} using the available photometry from Gaia DR2 (G/BP/RP), 2MASS (J/H/K$_{s}$), and {\it WISE} (W1/W2/W3). For the Gaia photometry, we assume the filter curves and zero points defined by \citet{Maiz2018} and photometric corrections published on the mission website\footnote{https://www.cosmos.esa.int/web/gaia/dr2-known-issues}. 

 
We find T$_{\rm eff} = 4239^{+84}_{-83}$, [Fe/H] = $+0.25^{+0.04}_{-0.05}$, $\rstar = 0.658^{+0.087}_{-0.088}$,  $\mstar = 0.703^{+0.079}_{-0.080}$, and $\lstar = 0.126 \pm 0.012$. These results from are consistent with those we get from \texttt{SpecMatch-Emp} and \texttt{isoclassify}. 
Our modeling suggests that \name has an enhanced alpha-element abundance ([$\alpha$/Fe]$\sim$+0.6). 

The \texttt{MINSWEEPER} analysis gives an age of 2.03$^{+3.77}_{-0.57}$ Gyr, which is inconsistent with the \texttt{isoclassify} analysis. However, we note that the age of low mass stars is very difficult to determine due to their long lifetimes and slow evolution on the main sequence. A more detailed analysis is needed to establish the age of this system but is beyond the scope of this study.

Due to the high SNR in the spectrum used to perform the spectroscopic analysis ($>350$ per resolution element) the formal statistical uncertainties are extremely small. We have seen from ensemble analysis of spectroscopic extraction codes and comparisons with other techniques that the statistical uncertainties from these codes are dwarfed by the systematic uncertainties in the stellar atmosphere and isochrone models in this high SNR regime \citep[e.g.,][]{Petigura17, Johnson17, Pepper19}.
For this reason, we follow the methodology of \cite{Petigura17} and \citet{Johnson17} by adding the following terms in quadrature with our measured \texttt{Specmatch-Emp}+\texttt{isoclassify} uncertainties; 100 K for T$_{\rm eff}$, 0.1 dex for $\log(g)$, and fractional uncertainties of 2\% for stellar mass, radius, and luminosity. These additional uncertainty terms were found by comparing the parameters estimated by \texttt{Specmatch-Emp}+\texttt{isoclassify} to those from other sources for a large sample of benchmark stars \citep{Yee17}. In addition, we attempt to incorporate the uncertainty due to our choice of modeling technique by adding the errors from \texttt{MINESweeper} in quadrature with those from \texttt{SpecMatch-Emp} and \texttt{isoclassify}. Our final adopted stellar parameters and associated uncertainties are listed in Table \ref{tab:MS}.

\subsection{Activity Indices}

As discussed earlier, \name is a relatively active K star.  We find that the median metrics of stellar activity as measured by the Ca II H\&K lines in our Keck/HIRES spectra are $\shk=0.98\pm0.11$ and $\lrphk=-4.72\pm0.05$ \citep{Isaacson10}. These values suggest a marginally consistent, but perhaps slightly lower activity level than the values reported in \citet{BoroSaikia18}, who reported $\shk=1.14$ and $\lrphk=-4.50$. The discrepancy between \lrphk\ from \citet{BoroSaikia18} and our value is due to differences in the methodology used to convert between \shk\ and \lrphk. If we convert their \shk\ value to \lrphk\ using the methodology of \citet{Isaacson10} we find a value of $\lrphk=-4.65$.

\begin{deluxetable}{lcc}
\tabletypesize{\footnotesize}
\tablewidth{0.4\textwidth}
\tablecaption{Stellar Properties of \name \label{tab:MS}}
\tablehead{
  \colhead{Parameter}   & 
  \colhead{\name} &
  \colhead{Source} 
}
\startdata
Alt. names               & HD 97101                   & \citet{Cannon93}\\
                         & HIP 54646                  & \citet{vanLeeuwen07}\\
RA                       & 11 11 05.17                & Gaia18\\
Dec                      & +30 26 45.66               & Gaia18\\
Spectral type            & K7V                        & \citet{Gray03} \\
$B-V$ (mag)              & 1.255                      & \citet{BoroSaikia18} \\
$m_V$ (mag)              & 8.864 $\pm$ 0.12           & \citet{Zacharias12} \\
$m_G$  (mag)             & 7.7281 $\pm$ 0.0007        & Gaia18\\
$m_J$  (mag)             & 5.764 $\pm$ 0.033          & \citet{Cutri03} \\
$m_H$  (mag)             & 5.130 $\pm$ 0.033          & \citet{Cutri03} \\
$m_K$  (mag)             & 4.979 $\pm$ 0.033          & \citet{Cutri03} \\
distance (pc)            & 11.893 $\pm$ 0.007         & Gaia18\\
T$_{\rm eff}$ (K)            & 4120 $\pm$ 109             & this work (\ref{SI}) \\
$\log(g)$                & 4.65 $\pm$ 0.04            & this work (\ref{MS})\\
{[Fe/H]}                 & 0.24 $\pm$ 0.10            & this work (\ref{SI}) \\
{[$\alpha$/Fe]}          & 0.58$^{+0.02}_{-0.08}$     & this work (\ref{MS}) \\
\rstar (\rsune)          & 0.680 $\pm$ 0.14           & this work (\ref{SI}) \\
\lstar (\lsune)          & 0.119 $\pm$ 0.013          & this work (\ref{SI}) \\  
\mstar (\msune)          & 0.650 $\pm$ 0.08         & this work (\ref{SI}) \\
Age (Gyr)                & 12.4 $\pm$ 5.2             & this work (\ref{SI}) \\
$A_v$ (mag)              & 0.02$^{+0.04}_{-0.01}$     & this work (\ref{MS}) \\
\cutinhead{ Gl 414B } 
RA                       & 11 11 02.54                & Gaia18 \\
Dec                      & +30 26 41.32               & Gaia18\\
Spectral type            & M2V                        & \citet{Stephenson86}  \\
$B-V$ (mag)              & 2.41 $\pm$ 0.34            &
\citet{Hog00}; \\
                         &                            & \citet{Zacharias12} \\
$m_V$ (mag)              & 9.983 $\pm$ 0.01           & \citet{Zacharias12}  \\
$m_G$  (mag)             & 9.0471 $\pm$ 0.0011        & Gaia18\\
$m_J$  (mag)             & 6.592 $\pm$ 0.019          &  \citet{Cutri03}\\
$m_H$  (mag)             & 5.975 $\pm$ 0.018          & \citet{Cutri03} \\
$m_K$  (mag)             & 5.734 $\pm$ 0.020          & \citet{Cutri03} \\
distance (pc)            & 11.877 $\pm$ 0.008         & Gaia18\\
T$_{\rm eff}$ (K)            & 3663 $\pm$ 70              & this work (\ref{SI})  \\
{[Fe/H]}                 & 0.08 $\pm$ 0.09            & this work (\ref{SI}) \\
\rstar (\rsune)          & 0.548 $\pm$ 0.017          & this work (\ref{SI}) \\
\lstar (\lsune)          & 0.048 $\pm$ 0.005          & this work (\ref{SI}) \\  
\mstar (\msune)          & 0.542 $\pm$ 0.022          & this work (\ref{SI}) \\
Age (Gyr)                & 11.2 $\pm$ 5.9 & this work (\ref{SI}) \\
\enddata
\end{deluxetable}
\vspace{20pt}

\section{Radial Velocity Measurements}
\label{sec:rv}

\subsection{Keck/HIRES}

We collected \nobskeck RV measurements between UT \kecksdate and \keckedate using the HIRES instrument on the Keck I telescope \citep{Vogt94}. This star was observed using the $0''.86\times3.5''$ and $0''.86\times14''$ slits for a spectral resolution of $R\approx65,000$ near 5500\AA. We extracted RV measurements from each spectrum following the technique described in \citet{Marcy_Butler92}. We observed \name with a cell of gaseous iodine in the light path just behind the entrance slit. The iodine cell imprints a dense forest of molecular absorption lines onto the stellar spectrum which are used as a simultaneous wavelength and point spread function (PSF) fiducial. We forward modeled each observation in 718 small chunks of spectral width $\approx$2 \AA\ as
\begin{equation}
    I_{obs}(\lambda) = k[T_{I_2}(\lambda) \cdot I_S(\lambda+\Delta\lambda)]\circledast \text{PSF},
\end{equation}
where $T_{I_2}$ is the transmission of the iodine cell as measured in a lab, $I_S(\lambda+\Delta\lambda)$ is the intrinsic stellar spectrum (perturbed by an RV that produces a wavelength shift of $\Delta\lambda$) derived by deconvolving an observed iodine-free stellar spectrum with the instrumental PSF. The product of $T_{I_2}$ and $I_S$ is then convolved with a model of the PSF which is described by a sum of Gaussians and scaled by an arbitrary normalization factor $k$. We list RVs and Ca II H \& K ($S_{HK}$) activity indices from Keck/HIRES in Table \ref{tab:rv}.

\subsection{APF/Levy}

We collected \nobsapf RV measurements between UT \apfsdate and \apfedate using the Levy spectrograph on the Automated Planet Finder (APF) telescope \citep{Vogt14, Radovan14}. We used the $1''\times3''$ slit for a spectral resolution of $R\approx$100,000 and reduced the data and extracted velocities in the same way as for Keck/HIRES. RVs and Ca II H \& K $S_{HK}$ activity indices from APF/Levy are also listed in Table \ref{tab:rv}.
  
\subsection{Shane/Hamilton Spectrograph}

This star was previously observed by the Twenty-five Year Lick Planet Search \citep{Fischer14}, with 13 RV measurements collected between UT Jan. 17, 1992 and Feb. 3, 2009 using the Hamilton Spectrograph on the Shane telescope at Lick Observatory. However, the scatter of these measurements is significantly higher than that of the other two datasets and we find that they add very little information to our fits. Including the Lick data would add several more free parameters to the fit, one for each of the CCD upgrades performed on the Hamilton Spectrograph. For this reason, we elect not to include these measurements in our final analysis.

\begin{deluxetable}{ccccc}
\tabletypesize{\footnotesize}
\tablecaption{\name Radial Velocities}
\tablewidth{245pt}
\tablehead{ 
    \colhead{\bjdtdb}               & \colhead{RV}  & \colhead{Uncertainty} & \colhead{Instrument\tablenotemark{1}} &  \colhead{\ensuremath{S_{\mbox{\scriptsize HK}}}} \\
    \colhead{(-- 2440000)}  & \colhead{(\mse)}            & \colhead{(\mse)}       & \colhead{} & \colhead{}
}
\startdata
10463.01045 & -1.39 & 1.31 & k & \nodata  \\
10546.933 & -20.30 & 1.16 & k & \nodata  \\
13370.0738 & 4.00 & 0.93 & j &  1.04 \\
13425.06406 & -4.16 & 0.98 & j & 1.01 \\
16589.03007 & -4.77 & 1.90 & a & 0.888 \\
16589.03878 & -3.88 & 1.76 & a & 0.888 \\
\enddata

\tablenotetext{}{(This table is available in its entirety in a machine-readable form in the online journal. A portion is shown here for guidance regarding its form and content.)}
\tablenotetext{1}{k = pre-upgrade Keck/HIRES, j = post-upgrade Keck/HIRES, a = APF}
\vspace{10pt}

\label{tab:rv}
\end{deluxetable}

\section{Visible-Light Photometry}
\label{sec:phot}

The Kilodegree Extremely Little Telescope \citep[KELT]{Pepper07} is  a small-aperture, wide-field photometric survey searching for transiting planets.  The KELT-North telescope is located at Winer Observatory, AZ,  and uses an Apogee AP16E detector with $4096\times4096$ 9$\mu$m pixels and a Mamiya 645 80mm lens. The KELT field of view is $26\degree \times 26\degree$, with $23''$ per pixel, and uses a non-standard wide bandpass comparable to a broad R-band filter.

KELT observed \name in KELT field KN07, obtaining 9516 photometric measurements between UT Dec. 24, 2006 and June 12, 2013. These data were reduced and photometry extracted following the procedures described in \cite{Siverd12}. We used the version of the light curve detrended with the TFA algorithm \cite{Kovacs05}. The KELT light curve for \name has an RMS of 7.3mmag and a median absolute deviation of 6.4mmag. With the large KELT pixels, the nearby companion Gl 414B is fully blended in the KELT aperture and contributes $\sim$20\% to the total flux.

\section{Doppler Analysis}
\label{sec:analysis}

\subsection{Search for Periodic Signals}

In order to identify significant periodic signals in the RVs, we used a hierarchical approach that compared a Keplerian model with a given number of planets to one with an additional planet to determine if adding a planet improved the quality of the fit at a statistically significant level. The Keplerian model is of the form:

\begin{align}
\label{eq:2}
    \mathcal{V}_r &= \sum_k^{N_{pl}} K_k[\cos(\nu_k + \omega_k) + e_k \cos(\omega_k)] + \gamma_i.
\end{align}
where $K$ is the velocity semi-amplitude, $e$ is the eccentricity, $\omega$ is the argument of periapsis, and $\nu$ is the true anomaly given by $\nu= 2\tan^{-1}\qty(\sqrt{\frac{1+e}{1-e}} \tan{\frac E 2})$, where $E$ is the eccentric anomaly. $\gamma_i$ is an offset term used to account for the different mean center-of-mass velocities of each instrument ($i$).  

We created a periodogram by using the Python package \texttt{RadVel} \citep{Fulton17}\footnote{\url{radvel.readthedocs.io}} version 1.2.13 to fit the RV data with a series of circular Keplerian models with fixed orbital periods. We utilized 10,000 search periods evenly spaced in frequency with values ranging between 20 and 6000 days. At each trial period we kept the period and eccentricity fixed while allowing all other orbital parameters to vary. We fit the model using the MAP optimization functionality built into \texttt{RadVel}. For fits with two or more planets we kept the orbital parameters of the previously-identified planet(s) fixed to the best-fit values from the $n-1$ planet version of the model. After completing our search in period space we represented the power at each trial period as the change in the Bayesian Information Criterion ($\Delta$BIC) \citep{schwarz1978} with respect to the $n-1$ planet model. 

\begin{figure}
    \centering
    \includegraphics[width=0.46\textwidth]{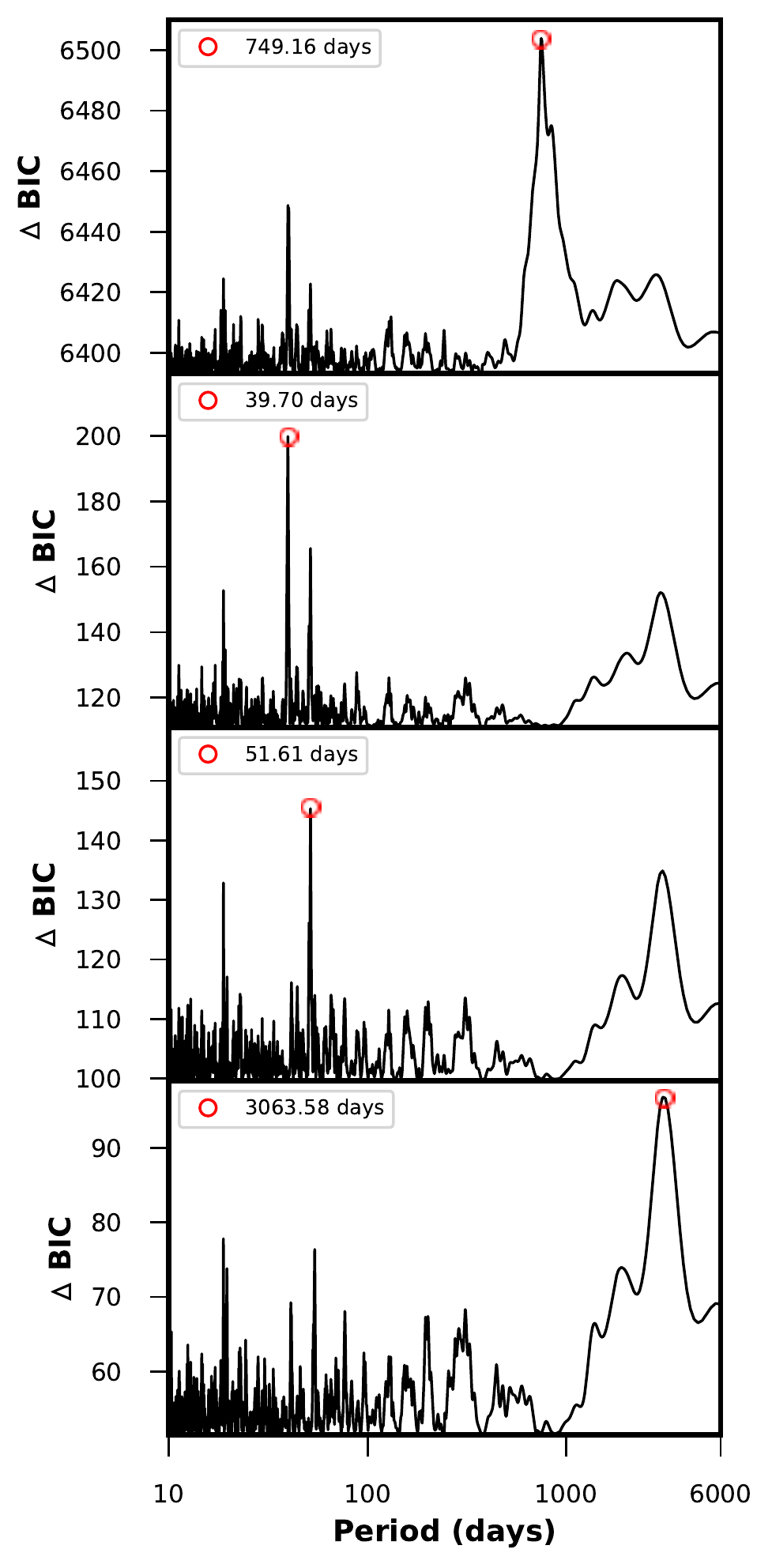}
    \caption{Results of the first four iterations of the multi-planet periodogram search described in Section 5.1. Shown here are the peaks at about 750, 40, 51, and 3000 days. Each panel has the previous highest-signal subtracted out to look for subsequent signals.}
    \label{fig:rv_pgrams}
\end{figure}

We began our search by comparing a model with one planet to a constant radial velocity  and found that the signal with the largest $\Delta$BIC was located at 749.16 days. We used this period as our initial guess and fit a new Keplerian orbit where the period and eccentricity were allowed to vary freely. We then incorporated this best-fit model into a new search for a second periodic signal. This revealed another peak at 39.7 days, which we fit with a model in which the parameters of both planets were allowed to vary. We then repeated this same process for two vs. three and three vs. four planets. We continued to iterate until the $\Delta$BIC of the highest peak was less than 10 relative to the noise and found significant periodic signals at 749.16, 39.70, 51.61, and 3063.58 days (Figure \ref{fig:rv_pgrams}).

\subsection{Stellar Activity}

\begin{figure*}[!htb]
    \centering
    \includegraphics{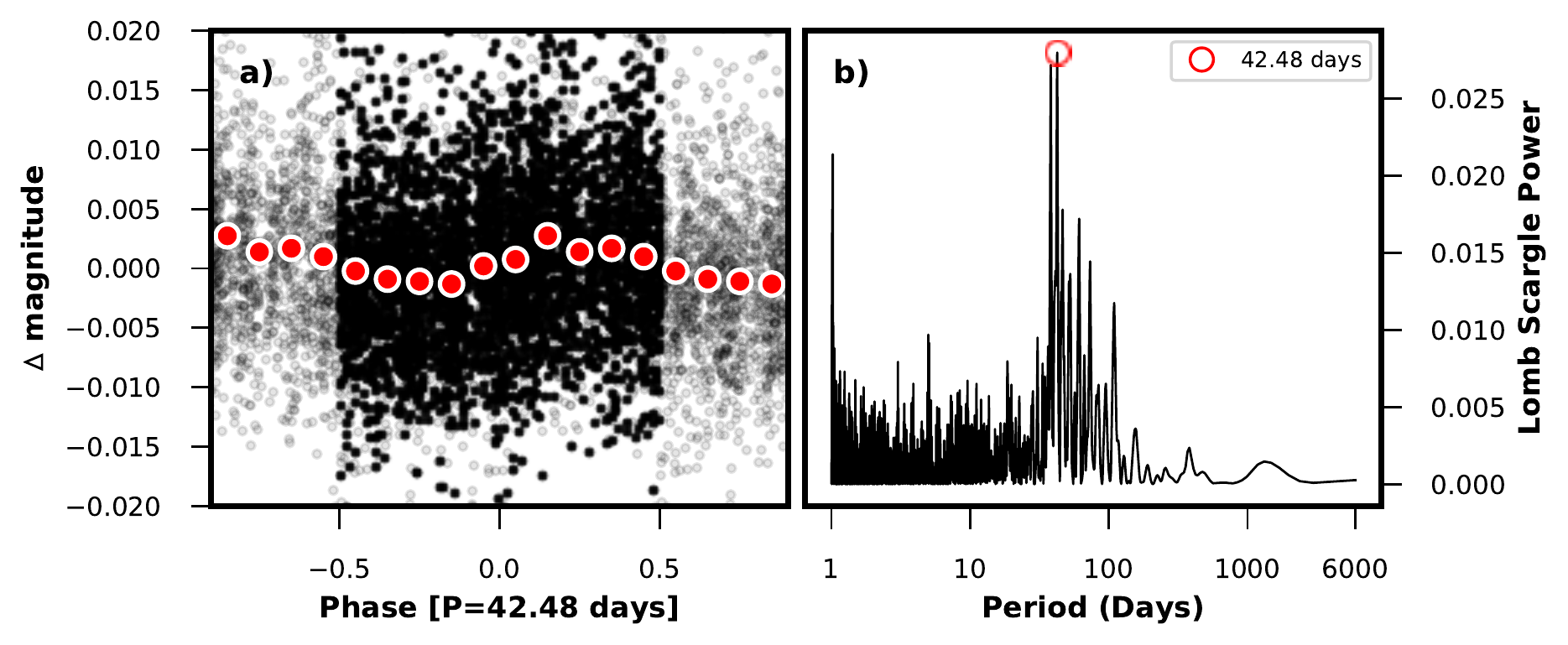}
    \caption{\textit{Left:}  KELT photometry phase-folded at a period of 42.48 days. The red dots show the values binned at intervals of 0.1 units of phase. \textit{Right:} Lomb-Scargle periodogram of the KELT photometry.}
    \label{fig:phot_pgram}
\end{figure*}

\begin{figure*}[!htb]
    \centering
    \includegraphics[width=\textwidth]{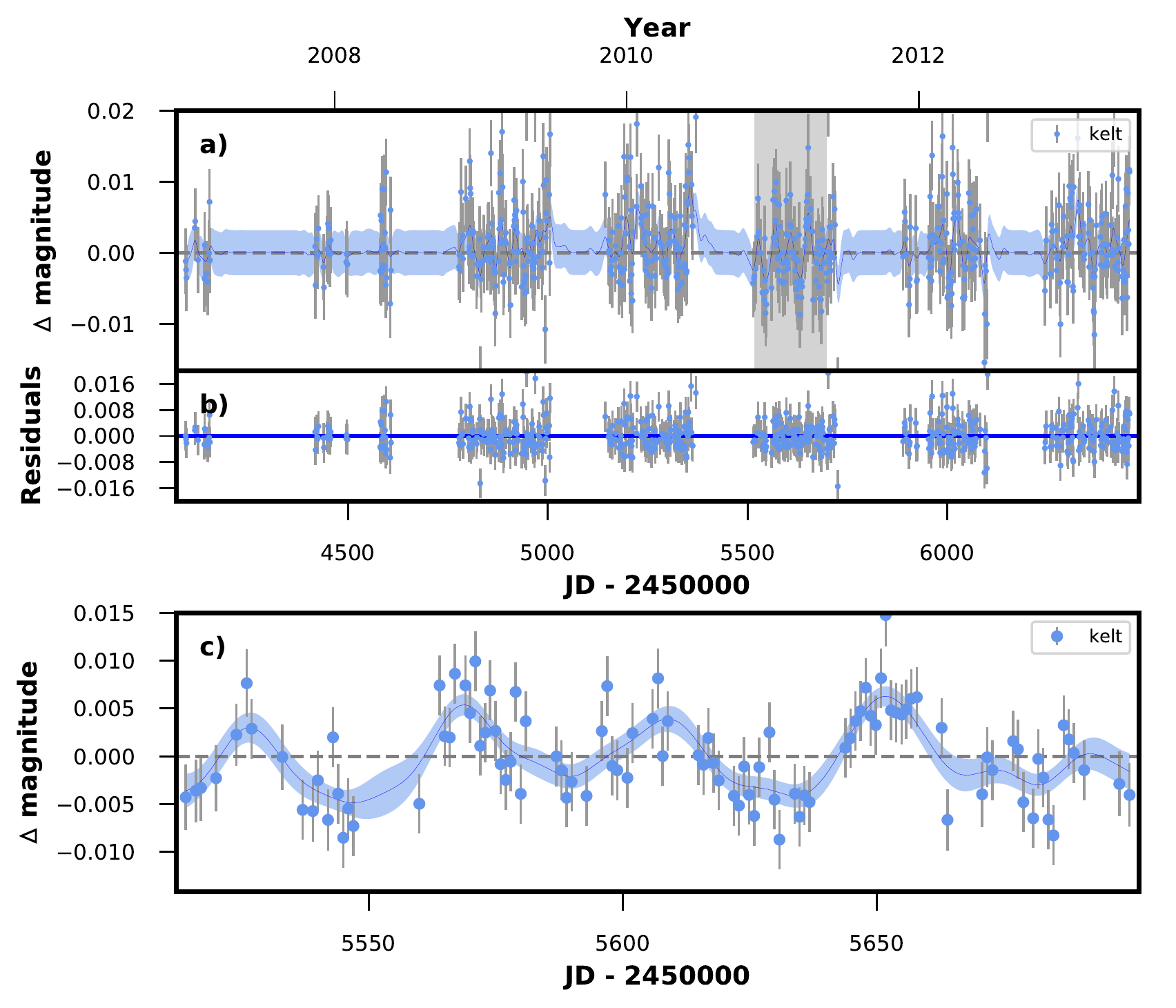}
    \caption{a) Photometric data from KELT (blue filled circles). The blue line shows the best fit Gaussian process model, with uncertainties shown as light blue bands. The grey highlighted section is displayed in more detail in panel c. b) Fit residuals c) A zoomed in view of a single season of photometric data where the quasi-periodic modulations are clearly visible.}
    \label{fig:phot_gp}
\end{figure*}

\begin{figure}[!htb]
    \includegraphics[width = 0.46\textwidth]{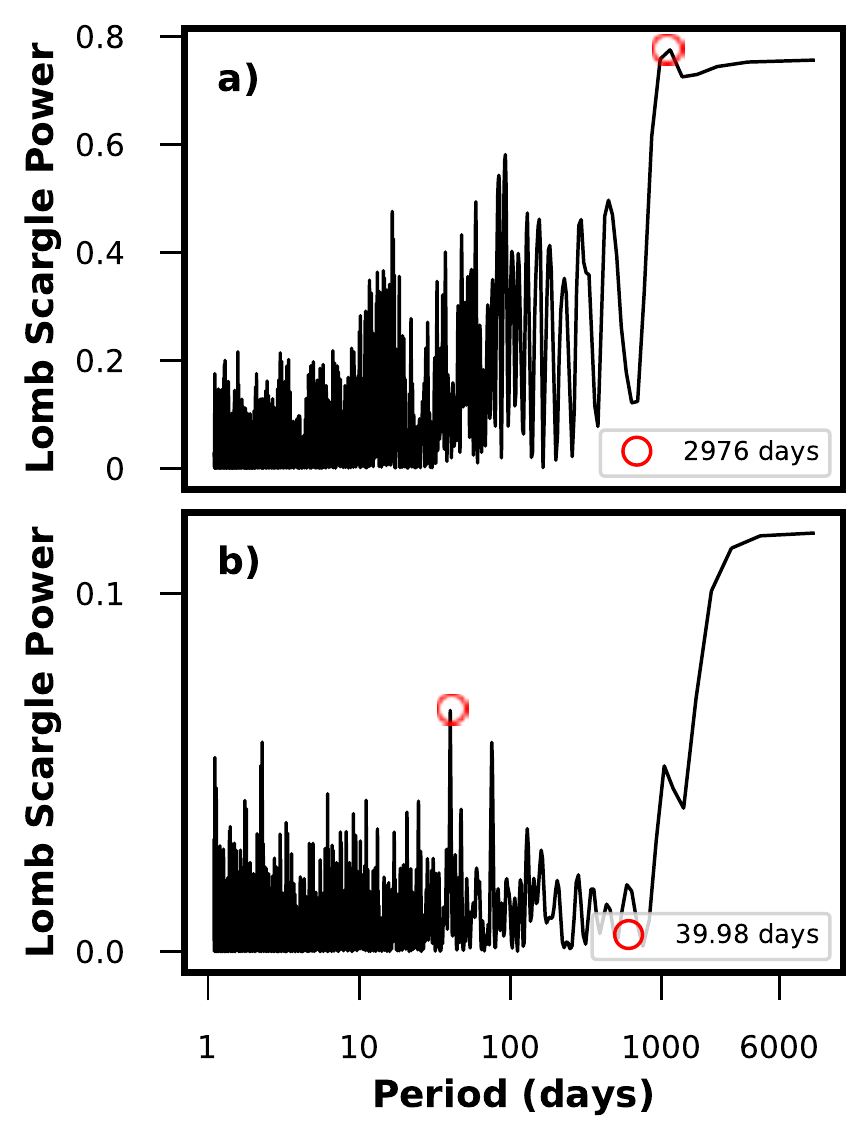}
    \caption{a) LS periodogram of the $S_{HK}$ values from Keck-HIRES. The peak seen here is related to the magnetic activity cycle of the star. b) LS periodogram of $S_{HK}$ values from APF/Levy. This signal is similar to the one we see at 39.6 days in the RVs and 42.5 days in the photometry.}
    \label{fig:svals}
\end{figure}

In order to determine whether or not any of the signals were related to stellar activity, we analyzed 6 years of photometric data from KELT. We used these data to create a Lomb-Scargle periodogram (Figure \ref{fig:phot_pgram}), which had a peak at a period of 42 days. We estimated the season-to-season variance in this period by splitting the data into individual observing seasons and identifying the highest peak for each season. We found that this variance was approximately 4 days.  We found an equivalent peak at 40 days in the $S_{HK}$ measurements of the star from APF (Figure \ref{fig:svals}). We therefore concluded that the 39.7 day signal in our RV periodograms was most likely a stellar activity signal related to the star's rotation period. 

The $S_{HK}$ periodogram from Keck had an additional strong peak at a period of 2976 days, which was similar to the RV periodogram peak at 3063 days in our four-planet model. We concluded that this signal was most likely related to the magnetic activity cycle of the star, which is typically on the order of a decade for FGK stars \citep{Lovis11}.

\subsection{Orbital Characterization}

We used the affine-invariant Markov Chain Monte Carlo (MCMC) sampler \texttt{emcee} \citep{Foreman-Mackey13} to determine the final posterior probability distribution for our two-planet model.  This functionality is available as a built-in feature of \texttt{RadVel}. We accounted for the approximately 39 day stellar activity signal using a Gaussian process (GP) model with a quasi-periodic kernel following a process similar to that described in \citet{Kosiarek19}. We chose not to include the 3063 day signal in this analysis because including it in the model does not significantly change the BIC. The elements of the covariance kernel are of the form:

\begin{align}
    C_{ij} &= \eta_1^2 \exp[-\frac{\abs{t_i-t_j}^2}{\eta_2^2}- \frac{\sin^2\qty(\frac{\pi \abs{t_i-t_j}}{\eta_3})}{2\eta_4^2}]
\end{align}
where the hyper-parameter $\eta_1$ is the amplitude, $\eta_2$ is the exponential decay timescale, $\eta_3$ is the period, and $\eta_4$ is the characteristic length of the periodic component.

The Keplerian and GP likelihoods are given by:

\begin{align}
    \ln \mathcal{L}_{i} &= - \frac 1 2 \sum_j \frac{(\mathcal{V}_{r,j} - d_j)^2}{e_j^2+ \sigma_i^2} - \ln \sqrt{2 \pi (e_j^2+ \sigma_i^2)} \\
    \ln \mathcal{L}_{gp} &= \frac12( r^T C^{-1} r - \ln[\det(C)] - N \ln(2 \pi)).
\end{align}
respectively, where $\mathcal{V}_r$ is the Keplerian model (\hyperref[eq:2]{Equation 2}), $e_j$ is the error associated with each data point $d_j$, $\sigma_i$ is the jitter term for each instrument, $r$ is the vector of residuals, $C$ is the covariance matrix, and $N$ is the number of measurements.

\begin{figure*}[!t]
    \centering
    \includegraphics[width=0.98\textwidth]{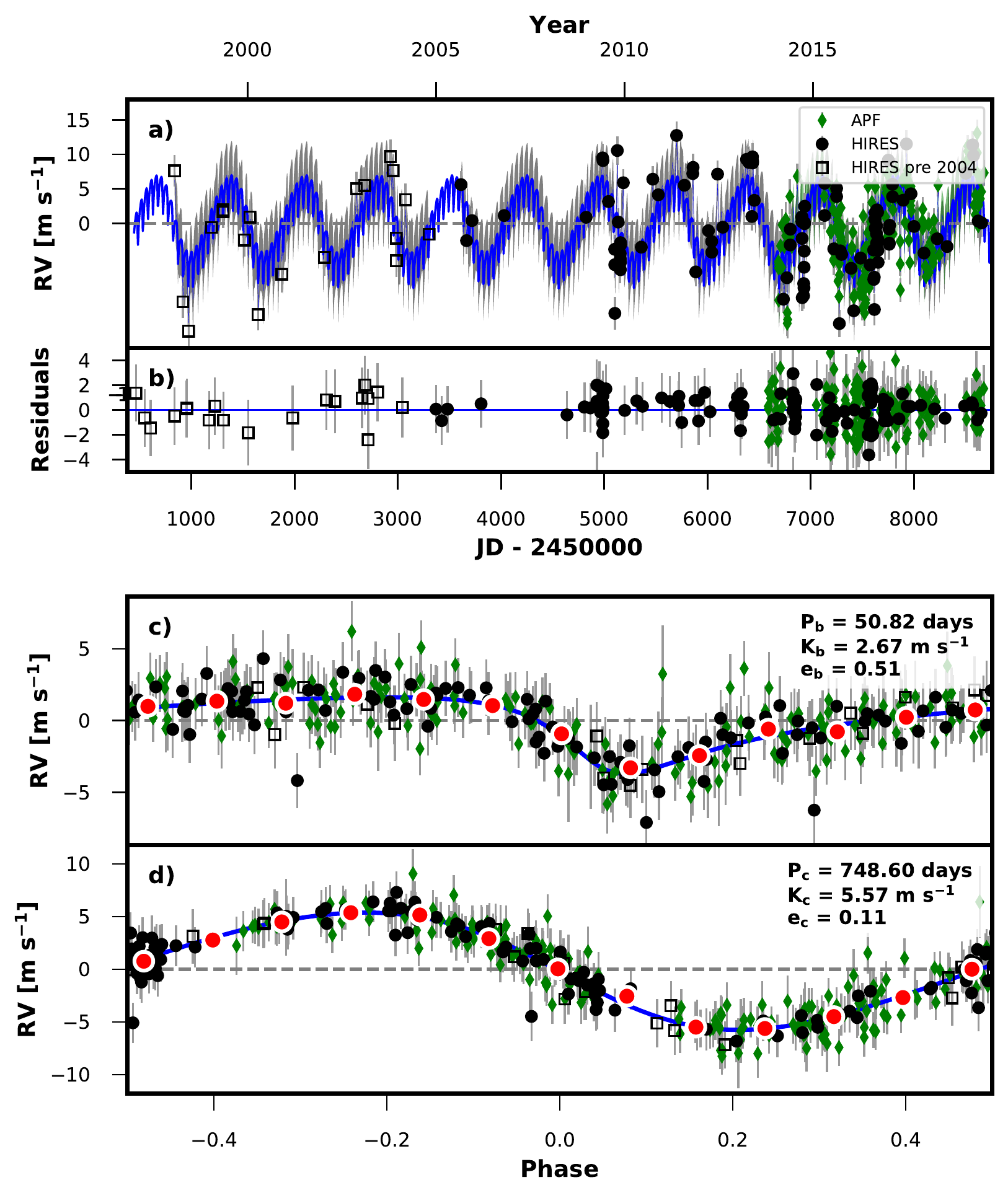}
    \caption{a) RV timeseries with the best fit two-planet model and Gaussian process stellar activity model overplotted in blue. The GP uncertainty is shown as grey bands. b) Fit residuals. c) RVs phase-folded to the ephemeris of planet b, with the Keplerian signal of planet c subtracted. Binned data is shown as red dots and the phase-folded model is shown as a blue line. d) RVs phase-folded to the ephemeris of planet c, with the Keplerian signal of planet b subtracted. Binned data are shown as red dots and the phase-folded model is plotted as a blue line.}
    \label{fig:rv_fit}
\end{figure*}

The total log-likelihood of all components of the model is then:

\begin{align}
    \ln \mathcal{L} = \sum_i \ln \mathcal{L}_i + \ln \mathcal{L}_{gp}. 
\end{align}
We first trained the GP on the photometric data to extract posteriors for the exponential decay length ($\eta_2$) and GP period ($\eta_3$). We then repeated this training on the $S_{HK}$ values using the posteriors from the photometric fit as priors. The priors are applied numerically using a kernel density estimator to approximate the shapes of the posteriors. The resulting posteriors on $\eta_2$ and GP period $\eta_3$ are then included as priors in the full RV fit. After training on both the photometric and $S_{HK}$ data sets, we found that $\eta_2 = 31.08\substack{+8.45 \\ -9.43}$ days and $\eta_3 = 40.11\substack{+6.34 \\ -4.00}$ days. We allowed $\eta_1$ to vary independently in each fit because the activity-driven amplitudes of the photometry, \shk values, and RVs are unrelated. We fixed the value of $\eta_4$ at 0.5, because allowing it to vary would cause unphysical effects in the fit, as shown in \cite{Lopez_Morales16}. The GP fit to the photometric data in Figure \ref{fig:phot_gp} indicates that the 40 day signal is not strictly periodic, as expected for stellar activity. We show the RV data and corresponding best-fit model in Figure \ref{fig:rv_fit}. 

\section{Discussion}
\label{sec:discussion}

We find compelling evidence for two planets in this system with orbital periods of 750 days and 51 days and minimum masses of $56.27$ M$_\Earth$ and $8.78$ M$_\Earth$, respectively.  We attribute the 40 day signal in the RV data to rotationally-modulated starspots, and the 3000 day signal to the magnetic activity cycle of the star. 

We used the Python code \texttt{Forecaster}, which is based on the probabilistic mass-radius relationship in \cite{Chen17}, to estimate radii corresponding to our minimum mass measurements. We predict that if its orbit is viewed close to edge-on, planet b should have a radius of $R_b =$ 2\rc \rearth and planet c should have a radius of $R_c =$ \rc \rearth. Assuming the bond albedo of planet b is equal to the mean total albedo of super-Earths \citep[$A_t = 0.32$,][]{Demory14} and  planet c is equal to that of Saturn \citep[$A_t = 0.343$,][]{2017Icar..282...19M}, we calculate predicted equilibrium temperatures of $T_{eq, b} = \Tb$ K and $T_{eq, c} = \Tc$ K for these two planets. We list the full set of derived properties for each planet in Table \ref{tab:orbital}.

\begin{deluxetable}{lrr}
\tabletypesize{\footnotesize}
\tablecaption{Planet Parameters}
\tablewidth{245pt}
\tablehead{ 
    \colhead{Parameter} & \colhead{Value} & \colhead{Units}
}
\startdata
  $P_{b}$              & $50.817 \substack{+0.031 \\ -0.03}$ & days \\
  $T\rm{conj}_{b}$     & $2454203.2 \substack{+1.9 \\ -2.4}$ & days \\
  $e_{b}$              & $0.48 \substack{+0.12 \\ -0.16}$    &  \\
  $\omega_{b}$         & $2.32 \substack{+0.42 \\ -0.59}$    & radians \\
  $K_{b}$              & $2.5 \substack{+0.58 \\ -0.65}$     & m s$^{-1}$ \\
  $e_b \sin{\omega_b}$ & $0.44 \substack{+ 0.19 \\ - 0.26}$  &  \\
  $e_b \cos{\omega_b}$ & $-0.46 \substack{+ 0.28 \\ - 0.20}$ &  \\
  $a_b$                & $\ab$                               & AU \\
  $M_b \sin{i}$        & \msinib                             & $M_\Earth$ \\
  $R_b$                &   \rb                               & $R_\Earth$ \\
    \vspace{10pt}
  $T_{eq, b}$          & $\Tb$                               & K \\
  $P_{c}$              & $748.3 \substack{+1.3 \\ -1.2}$     & days \\
  $T\rm{conj}_{c}$     & $2454205.0 \substack{+5.0 \\ -5.4}$ & days \\
  $e_{c}$              & $0.093 \substack{+0.1 \\ -0.064}$   &  \\
  $\omega_{c}$         & $1.6 \substack{+0.74 \\ -2.8}$      & radians \\
  $K_{c}$              & $5.45 \substack{+0.77 \\ -0.75}$    & m s$^{-1}$ \\
  $e_c \sin{\omega_c}$ & $0.21 \substack{+ 0.18 \\ - 0.16}$  &  \\
  $e_c \cos{\omega_c}$ & $-0.07 \substack{+ 0.19\\ - 0.26}$  &  \\
  $a_c$                & $\ac$                               & AU \\
  $M_c \sin{i}$        & \msinic                             & $M_\Earth$ \\
  $R_c$                &  \rc                                & $R_\Earth$ \\
  $T_{eq, c}$          & $\Tc$                                 & K \\
\enddata

\vspace{10pt}

\label{tab:orbital}
\end{deluxetable}

\subsection{Orbital Dynamics}

\begin{figure}[!htb]
    \includegraphics[width = 0.46\textwidth]{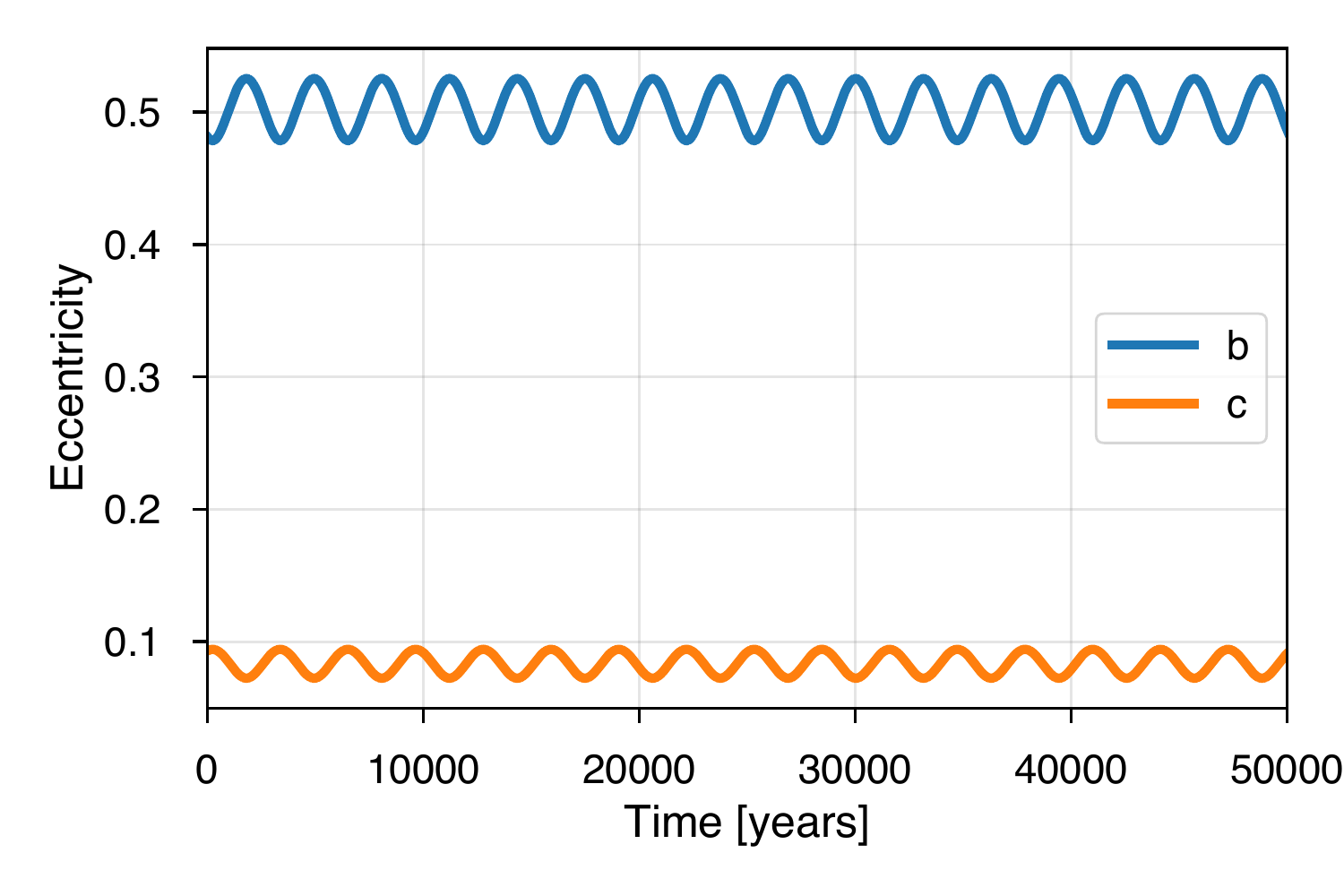}
    \includegraphics[width = 0.46\textwidth]{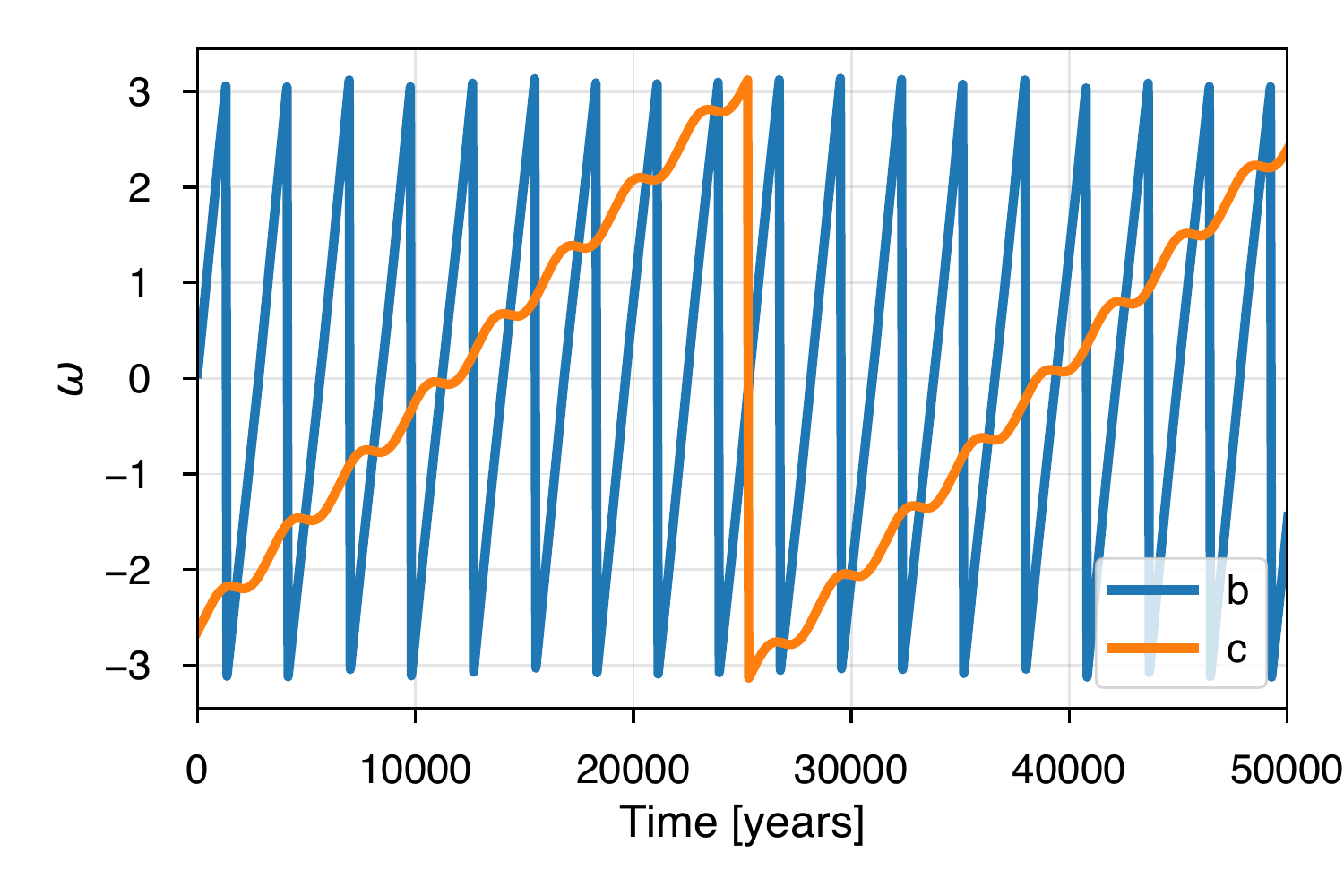}
    
    \caption{Evolution of orbital parameters from a N-body simulation assuming coplanar orbits. \emph{Top:} Eccentricity as a function of time for each planet.
    \emph{Bottom:} Argument of periastron ($\omega$) for each planet as a function of time. Each planet's eccentricity oscillates slightly as $\omega_b$ precesses with a timescale of $\approx$2.5 kyrs. $\omega_c$ preccesses more slowly, with a timescale of $\approx25$ kyrs.}
    \label{fig:rebound_sim}
\end{figure}

We simulated the system using the \texttt{REBOUND} N-body integration code \citep{Rein12, Rein15} to check for stability and investigate the orbital dynamics. We initially assumed that the orbits are coplanar and set the orbital elements to their MAP values. We simulated the system in this configuration for 10 million years with a time step of 0.17 days. The system remained stable in this configuration throughout the simulation. Figure \ref{fig:rebound_sim} shows the evolution of eccentricity and the argument of periastron ($\omega$) as a function of time for the first 50 thousand years of the simulation. $\omega_{b}$ precesses at a relativley rapid rate of 360 degrees every $\sim$2500 years. The eccentricities of both planets trade off with each other on this same timescale. $\omega_c$ precesses more slowly, with a period of $\sim$25000 years, but the rate of precession oscillates at the shorter timescale ($\sim$2500 years). To test the sensitivity of these precession rates to the initial conditions we also draw 10 sets of orbital parameters from the posterior distributions as starting points for a suite of N-body simulations. We find that the precession timescales vary by $\sim30\%$, but the general properties of the orbital evolution remain constant.

\begin{figure}[!htb]
    \includegraphics[width = 0.5\textwidth]{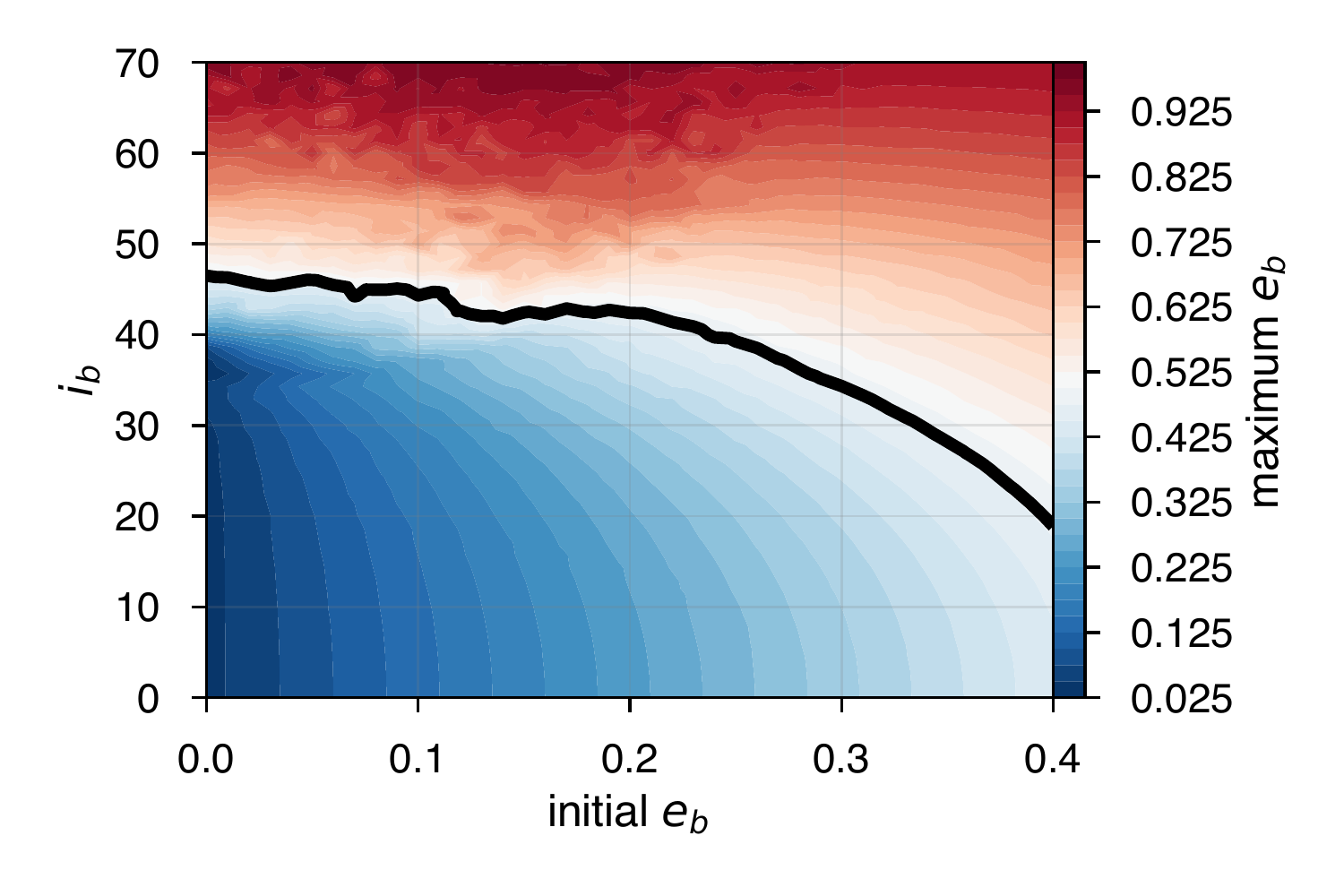}
    \caption{Grid of N-body simulations. Each simulation starts with a different initial eccentricity and inclination for the inner planet. The color-scale shows the maximum eccentricity reached by the inner planet during each 50 kyr simulation. The thick black lines shows the minimum combination of initial eccentricity and mutual inclination needed to explain the current eccentricity of planet b.}
    \label{fig:sim_grid}
\end{figure}

We measure a significant eccentricity for the inner planet ($e_b=0.48^{+0.12}_{-0.16}$) while the outer, more massive planet, is on a nearly circular orbit. The relatively large eccentricity of the inner planet in this system might plausibly be due to some kind of past dynamical instability (Huang et al. 2017, Carrera et al. 2019).  If the two planets have a mutual inclination greater than $\sim$45$\degree$, the inner planet could also be due to Kozai-Lidov oscillations where it cycles between states with high inclination and high eccentricity \citep{Kiseleva98}. However, for planets close to their host stars general-relativistic (GR) precession may also be significant and can dampen the amplitude of the Kozai-Lidov oscillations \citep[e.g.,][]{Ford00, Fabrycky07}. Following the methodology of \cite{Yee18} we calculate the GR and Kozai timescales for planet b. We find that the GR timescale is $\tau_{\rm GR}\sim5$ Myrs and the Kozai timescale is $\tau_{\rm Kozai}\sim24$ kyrs. Since $\tau_{\rm Kozai} \ll \tau_{\rm GR}$ it is possible that this system is currently undergoing Kozai oscillations.

We explored the impact of mutual inclination on the system architecture by running many \texttt{REBOUND} simulations, each time perturbing the inclination and initial eccentricity of the inner planet keeping the other orbital elements fixed to their MAP values. For each simulation we recorded the maximum eccentricity reached by the inner planet during the 50 thousand year simulation (Figure \ref{fig:sim_grid}). We find that if the orbit of the inner planet was initially circular, the mutual inclination of the system must be $\gtrsim$45 degrees in order to explain the planet b's current eccentricity.

For planets that come close to the star during periastron, tidal circularization is expected to damp out the eccentricity induced by Kozai osciallations or other dynamical interactions (e.g., Fabrycky \& Tremaine 2007).  We estimated the tidal circularization timescales for this system using the method outlined in \cite{Albrecht12}. Planets b and c both have extremely long tidal circularization times, at $3.2 \times 10^{11}$ Gyr and $3.8 \times 10^{14}$ Gyr respectively.  We therefore expect that any dynamically-induced orbital eccentricity should persist to the present day.

\subsection{Transit Search}

\begin{figure}[!htb]
    \centering
    \includegraphics[width=0.46\textwidth]{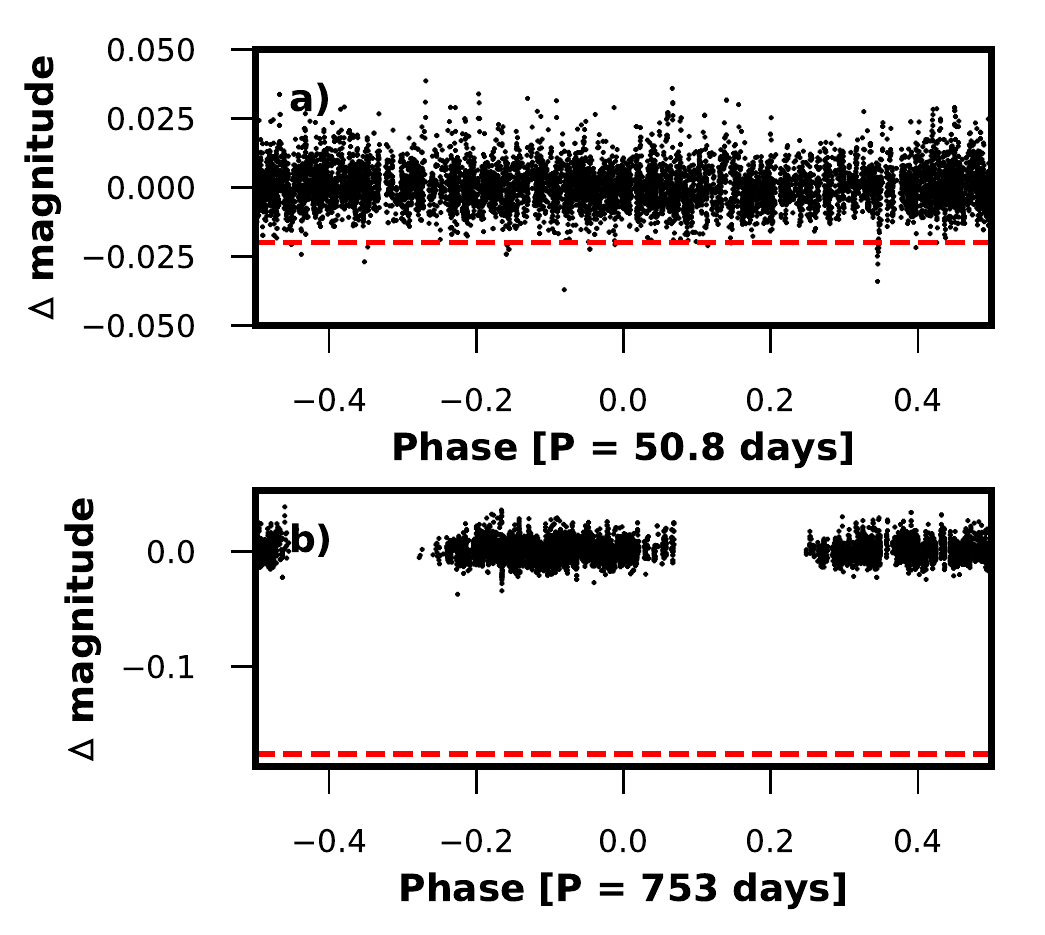}
    \caption{KELT photometry phase-folded on the ephemeris of planet b (top) and planet c (bottom). The dotted red line shows the predicted transit depth. We see no transit signals at these depths and locations}
    \label{fig:transit}
\end{figure}

We calculate a priori transit probabilities of 1\% and 0.2\% for planets b and c, respectively. The a posteriori transit probabilities are likely slightly higher \citep{Stevens13}, but it is unlikely that either of them transits. If they were to transit, we can use the radius estimates from the previous section to calculate predicted transit depths of 0.16\% for planet b and 1.4\% for planet c after accounting for 20\% dilution in the KELT photometry due to Gl 414B. We plot the KELT photometry phase-folded on the ephemeris of each planet in Figure \ref{fig:transit}. We note that there appears to be what looks like a transit signal in both phase-folded light curves, but that the phase of these two signals is inconsistent with the predicted time of conjunction for each of the two planets. Upon closer inspection, both signals also occur adjacent to a gap in the data, where instrumental and telluric variations are more likely to create a false positive.  We therefore conclude that there is no evidence to suggest that either planet is transiting.

\subsection{Direct Imaging Prospects}

At a distance of just 12 pc, this relatively cool star is a promising prospect for direct imaging studies. We find that the planets in this system have projected separations of 20 and 120 mas. 

We first compare these separations to the inner working angle (IWA) of two ground-based direct imaging instruments. The Keck Planet Imager and Characterizer (KPIC) \citep{Mawet18} has an IWA of 40-100 mas, and Spectro-Polarimetric High contrast imager for Exoplanets REsearch (SPHERE) \citep{2019arXiv190204080B} has an IWA of 50-80 mas. Gl 414c is beyond the IWA of all of these instruments. However, in the K band, the contrast would be $2.5 \times 10^{-11}$, which rules out ground-based instruments that operate in the near-infrared, like KPIC and SPHERE.

We also compare these angles to the IWA of JWST coronographs. The only mask with an IWA less than the separation of Gl 414c is the non-redundant mask (NRM) on the Near Infrared Imager and Slitless Spectrograph (NIRISS). The minimum IWA of the NRM is 89 mas, when observing with the F277W filter. At 2.77 $\mu$m, the contrast of planet c with respect to its host star is $1.27 \times 10^{-7}$. At this contrast, Gl 414c is not visible to NIRISS \citep{Doyon12}. 

However, given its contrast in the mid-IR ($10^{-6}$ at 10 $mu$m) this planet may be a candidate in the future for imaging with a mid-infrared adaptive optics system on a thirty meter-class telescope \citep{chun06}.

\subsection{Habitability}

Using the model described in \cite{Kopparapu13}, we calculate that the habitable zone around this star should lie between 0.37 and 0.70 AU. The planet c falls well outside this zone, but planet b is just inside this inner edge. If we use the more optimistic model from \cite{2013ApJ...778..109Z}, the inner edge of the habitable zone may be as close as 0.21 AU. With a semi-major axis of 0.24 AU, planet b would fall within this habitable zone range. However, planet b has a minimum mass of $8.8$ \mearth which likely corresponds to a substantial volatile-rich envelope \citep{weiss14}, so it is not a good candidate for habitability. 

\section{Summary \& Conclusion}
\label{sec:summary}

\begin{figure*}[!htb]
    \centering
    \includegraphics[width=\textwidth]{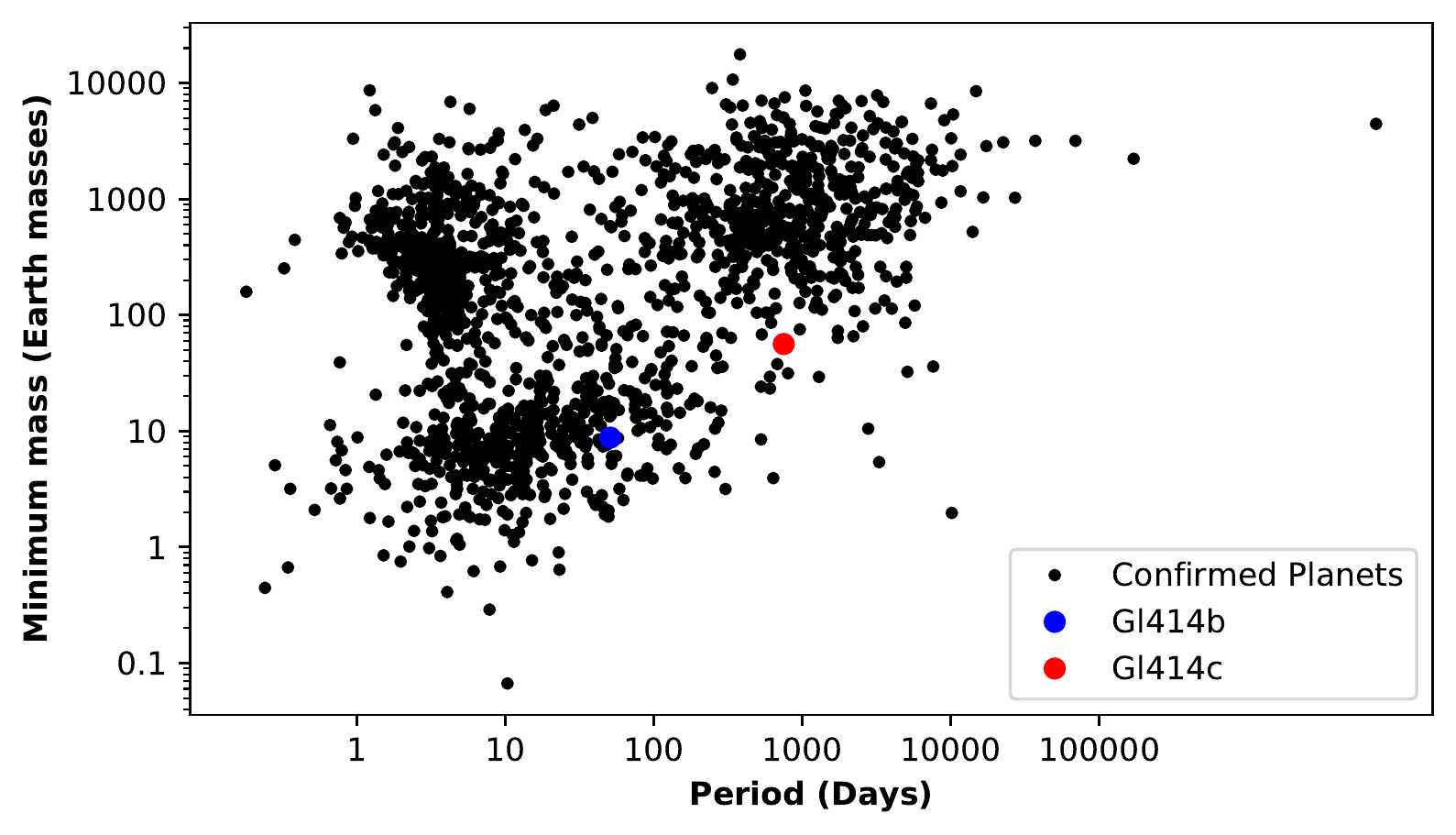}
    \caption{Plot of $M \sin i$ vs. orbital period for all known exoplanets as of 8/28/2020 with measured masses, with \name b and \name c placed in context. This plot illustrates that these two planets are among the longest period planets that have been detected to date in their respective mass ranges.}
    \label{fig:all}
\end{figure*}

We present the discovery of a sub-Neptune planet and a sub-Saturn planet orbiting the bright K7 dwarf \ename. The minimum masses of the planets are $M_b \sin i_b =$ \msinib $M_\Earth$ and $M_c \sin i_b =$ \msinic $M_\Earth$ and they orbit with semi-major axes $a_b = \ab$ AU and $a_c = \ac$ AU. Planet c resides near the inner edge of the star's habitable zone, but its minimum mass is large enough that it likely possesses a substantial volatile-rich envelope. Figure \ref{fig:all} shows the period vs. mass of all currently known planets detected using the radial velocity technique and places Gl 414b and Gl 414c in context. 

In a search of the 4201 confirmed planets listed in the NASA Exoplanet Archive, we found eight other planets in multi-planet systems where one of the planets had an orbital period between 30 and 100 days and an eccentricity greater than 0.4. (HD 163607, Kepler-419, HD 168443, HD 37605, Kepler-432, HD 74156, V1298 Tau, and HD 147018.)\footnote{\url{https://exoplanetarchive.ipac.caltech.edu/}} In seven of these systems, there are exactly two planets detected. The high eccentricity planet is always the inner planet, and the outer planet always has a large separation from the inner planet ($P > 400$ days). This suggests that this class of planets may have a common dynamical origin; for example, it is possible that these system architectures all arose from Kozai-Lidov oscillations. These systems may represent a dynamically active subset of the larger exoplanet population.

\acknowledgments

We are grateful to the time assignment committees of the California Institute of Technology, the University of California, the University of Hawai‘i, and NASA for their generous allocations of observing time. Without their long-term commitment to RV monitoring, these planets would likely remain unknown.  We gratefully acknowledge the efforts and dedication of the staffs of the W.\ M.\ Keck Observatory and Mt.\ Hamilton. Research at the Lick Observatory is partially supported by a generous gift from Google.  We thank Geoff Marcy and Debra Fischer for their many nights of observing that contributed to the Keck data presented in this work, and R.\ Paul Butler and S.\ S.\ Vogt for many years of contributing to the data presented here. AWH acknowledges NSF grant AST-1517655. MRK acknowledges support from the NSF Graduate Research Fellowship, grant No. DGE 1339067. DJS was supported as an Eberly Research Fellow by the Eberly College of Science at the Pennsylvania State University.

The authors extend special thanks to those of Hawaiian ancestry on whose sacred mountain of Maunakea we are privileged to be guests. Without their generous hospitality, the Keck observations presented herein would not have been possible. 

This work made use of the GNU Parallel package for large-scale multiprocessing \citep{OTange18}. This research has made use of the NASA Exoplanet Archive, which is operated by the California Institute of Technology, under contract with the National Aeronautics and Space Administration under the Exoplanet Exploration Program. This work made extensive use of the scipy \citep{Virtanen19}, numpy \citep{Vanderwalt11}, matplotlib \citep{Hunter07}, and pandas \citep{Mckinney10} Python packages.

Simulations in this paper made use of the \texttt{REBOUND} code which is freely available at \url{http://github.com/hannorein/rebound}.

\facility{Automated Planet Finder (Levy), Keck:I (HIRES), Exoplanet Archive}

\break

\bibliographystyle{aasjournal}
\bibliography{references}

\end{document}